\documentclass[12pt]{article}
\usepackage{graphicx}
\usepackage{epsfig}
\usepackage{epstopdf}
\DeclareGraphicsExtensions{.pdf,.eps,.png,.jpg,.mps}

\def\lsim{\mathrel{\rlap{\lower3pt\hbox{\hskip0pt$\sim$}}
     \raise1pt\hbox{$<$}}}         
\def\gsim{\mathrel{\rlap{\lower4pt\hbox{\hskip1pt$\sim$}}
     \raise1pt\hbox{$>$}}}         

\usepackage{amsmath}
\usepackage{amsfonts}

\begin{document}
\begin{titlepage}

\centerline{\Large \bf Combining Alpha Streams with Costs}
\medskip

\centerline{Zura Kakushadze$^\S$$^\dag$\footnote{\, \tt Email: zura@quantigic.com}}

\bigskip

\centerline{\em $^\S$ Quantigic$^\circledR$ Solutions LLC}
\centerline{\em 1127 High Ridge Road \#135, Stamford, CT 06905\,\,\footnote{\, DISCLAIMER: This address is used by the corresponding author for no
purpose other than to indicate his professional affiliation as is customary in
publications. In particular, the contents of this paper
are not intended as an investment, legal, tax or any other such advice,
and in no way represent views of Quantigic® Solutions LLC,
the website \underline{www.quantigic.com} or any of their other affiliates.
}}
\centerline{\em $^\dag$ Department of Physics, University of Connecticut}
\centerline{\em 1 University Place, Stamford, CT 06901}
\medskip
\centerline{(May 19, 2014; revised: July 7, 2014)}

\bigskip
\medskip

\begin{abstract}
{}We discuss investment allocation to multiple alpha streams traded on the same execution platform with internal crossing of trades and point out differences with allocating investment when alpha streams are traded on separate execution platforms with no crossing. First, in the latter case allocation weights are non-negative, while in the former case they can be negative. Second, the effects of both linear and nonlinear (impact) costs are different in these two cases due to turnover reduction when the trades are crossed. Third, the turnover reduction depends on the universe of traded alpha streams, so if some alpha streams have zero allocations, turnover reduction needs to be recomputed, hence an iterative procedure. We discuss an algorithm for finding allocation weights with crossing and linear costs. We also discuss a simple approximation when nonlinear costs are added, making the allocation problem tractable while still capturing nonlinear portfolio capacity bound effects. We also define ``regression with costs" as a limit of optimization with costs, useful in often-occurring cases with singular alpha covariance matrix.
\end{abstract}

\bigskip
\medskip

{}{\bf Keywords:} hedge fund, alpha stream, crossing trades, transaction costs, impact, portfolio turnover, investment allocation, weight optimization

\end{titlepage}

\newpage

\section{Motivation and Summary}

{}Combining multiple hedge fund alpha streams has the benefit of diversification.\footnote{\, For a partial list of hedge fund literature, see, {\em e.g.}, \cite{HF1}-\cite{HF20} and references therein.} One then needs to determine how to allocate investment into these different alpha streams $\alpha_i$, or, mathematically speaking, how to determine the weights $w_i$ with which the investment should be allocated to individual alphas.\footnote{\, By ``alpha" we mean any ``expected return". {\em A priori} it need not even be stock based.}

{}If individual alpha streams are traded on separate execution platforms, then the weights are non-negative: $w_i \geq 0$. This applies to the hedge fund of funds vehicles, which take long positions in individual hedge fund alpha streams, as well as long-only mutual funds. Also, this is irrespective of whether transaction costs are included or not. So, the investment allocation problem then is some portfolio optimization problem whereby one determines the weights $w_i$ based on an optimization criterion\footnote{\, For a partial list of portfolio optimization and related literature, see, {\em e.g.}, \cite{PO1}-\cite{PO35} and references therein.} -- {\em e.g.}, maximizing the Sharpe ratio, the P\&L, maximizing the P\&L subject to a condition on the Sharpe ratio, {\em etc.} -- and invariably this portfolio optimization involves the requirement that the weights are non-negative. Since the weights are non-negative, if we include linear costs $L_i$, the P\&L is simply given by ($I$ is the investment level; see Section \ref{sub2.1} for more detail):
\begin{equation}
 P = I \sum_i \left(\alpha_i - L_i\right)~w_i
\end{equation}
So adding the linear cost $L_i>0$ simply has the effect of reducing the alpha $\alpha_i$.

{}Combining and trading multiple hedge fund alpha streams on the same execution platform has a further benefit that by internally crossing the trades between different alpha streams (as opposed to going to the market) one benefits from substantial savings on transaction costs.\footnote{\, For a recent discussion, see \cite{OD}.} In this framework the weights with which the alphas are combined need no longer be non-negative. This is because, due to different alphas being correlated with each other, the optimal allocation for the wights can be such that some alphas are traded in reverse, against their originally intended signal.\footnote{\label{neg.w}\, {\em E.g.}, consider two alphas $\alpha_1>0$ and $\alpha_2>0$ with unit variances and correlation $\rho>0$, with no costs. The Sharpe ratio $S$ is maximized by $w_1 = \gamma(\alpha_1 - \rho~\alpha_2)$, $w_2 = \gamma(\alpha_2 - \rho~\alpha_1)$, where $\gamma$ is fixed from $|w_1| + |w_2| = 1$. If $\alpha_2 < \rho~\alpha_1$, then $w_2 < 0$, so $S\rightarrow\mbox{max}$ requires ``shorting" $\alpha_2$.} On the one hand, we no longer have the $w_i \geq 0$ bound, which simplifies the optimization problem. On the other hand, when costs are included, this leads to a complication, because the costs are positive whether a given alpha is traded along or against the signal. {\em E.g.}, in the case of linear costs, the P\&L now becomes (assuming for the sake of simplicity the same linear cost regardless of the direction of trading; see Section \ref{sub2.1} for more detail)
\begin{equation}
 P = I \sum_i \left(\alpha_i~w_i - L_i~\left|w_i\right|\right)
\end{equation}
It is the modulus in $L_i~\left|w_i\right|$ that complicates the weight optimization problem, both in the case of linear costs only, as well as when nonlinear costs -- or impact of trading on prices -- are included.

{}Yet another issue arises when one accounts for turnover reduction due to internal crossing. Without internal crossing, turnover $T$ of the combined portfolio is simply the weighted sum of the individual turnovers $\tau_i$ (by $\tau_i \equiv D_i / I_i$ we mean the percentage of the dollar turnover $D_i$ of the individual alpha stream $\alpha_i$ with respect to the total dollar investment $I_i$ into this alpha stream assuming it is traded separately, without any crossing with other alpha streams):
\begin{equation}
 T = \sum_i \tau_i \left|w_i\right|
\end{equation}
However, when trades are crossed, turnover reduces,\footnote{In this regard, optimizing alpha streams in the context of trading them on the same execution platform is different from stock portfolio optimization. With stocks, there is no ``internal crossing" or ``turnover reduction". Yet another difference is that with stocks, for dollar-neutral portfolios, there is a constraint $\sum_a d_a = 0$, where $d_a$ is the dollar holding for the stock labeled by $a$. For alphas we have the condition on the weights $\sum_i |w_i| = 1$ instead.}  and when the number of alphas is large, the following model is expected to provide a good approximation \cite{SpMod}:
\begin{equation}
 T \approx \rho_* \sum_i \tau_i \left|w_i\right|
\end{equation}
where $0 <\rho_*\leq 1$ is the turnover reduction coefficient. In \cite{SpMod} we proposed a spectral model for estimating $\rho_*$, which is based on the correlation matrix of the alphas and is designed to work when the number of alphas is large, and the distribution of individual turnovers $\tau_i$ is not skewed. The turnover reduction coefficient $\rho_*$ depends on the universe of alphas that are being traded -- this is the case in the aforementioned spectral model, and is also expected to be a model-independent property. In this regard, if upon solving the optimization problem some weights $w_i$ turn out to be zero, then $\rho_*$ needs to be recomputed with the corresponding alphas dropped, and optimization needs to be repeated with so recomputed $\rho_*$. In fact, this process needs to be repeated iteratively until it converges. This is yet another feature specific to alpha stream optimization with internal crossing.

{}Thus, determining the optimal allocation of weights when alphas are traded on the same execution platform and trades are internally crossed is a rather different optimization problem from finding the weights when one combines alpha streams traded on separate trading platforms. In this note, motivated by these differences, we discuss the optimization problem in the framework of combining alphas traded on the same trading platform with internal crossing. We discuss an algorithm (which requires a finite number of iterations) for finding $w_i$ in the presence of linear costs. The optimization criterion is taken to be maximizing the Sharpe ratio, and the alpha covariance matrix is taken to be of a factor model form. We also discuss the case when nonlinear costs (impact) are added. We discuss a simple approximation in this case which makes the optimization problem tractable while still capturing the nonlinear dependence on the investment level that governs the portfolio capacity bounds. We also discuss the case where the alpha covariance matrix is singular, which often occurs in practical applications, in which case we discuss how to do ``regression with costs", which is a limit of optimization with costs.

{}The remainder of this paper is organized as follows. In Section \ref{sec.setup} we give our notations and setup. In Section \ref{sec.opt} we discuss optimization with linear costs. In Section \ref{sec.impact} we discuss optimization with linear costs plus impact. In Section \ref{sec.reg} we discuss the regression limit of optimization with costs.

\section{Definitions and Setup}\label{sec.setup}

{}We have $N$ alphas $\alpha_i$, $i=1,\dots,N$. Each alpha is actually a time series $\alpha_i(t_s)$, $s=0,1,\dots,M$, where $t_0$ is the most recent time. Below $\alpha_i$ refers to $\alpha_i(t_0)$.

{}Let $C_{ij}$ be the covariance matrix of the $N$ time series $\alpha_i(t_s)$. Let $\Psi_{ij}$ be the corresponding correlation matrix, {\em i.e.},
\begin{equation}
 C_{ij} = \sigma_i~\sigma_j~\Psi_{ij}
\end{equation}
where $\Psi_{ii} = 1$. If $M < N$, then only $M$ eigenvalues of $C_{ij}$ are non-zero, while the remainder have ``small" values, which can be positive or negative. These small values are zeros distorted by computational rounding.\footnote{\, Actually, this assumes that there are no N/As in any of the alpha time series. If some or all alpha time series contain N/As in  non-uniform manner and the correlation matrix is computed by omitting such pair-wise N/As, then the resulting correlation matrix may have negative eigenvalues that are not ``small" in the sense used above, {\em i.e.}, they are not zeros distorted by computational rounding. The deformation method mentioned below can be applied in this case as well.\label{corrNAs}} In such cases, one can deform the covariance matrix so it is positive-definite (see Subsection 3.1 of \cite{SpMod} for a deformation method based on \cite{RJ}). Still, the off-diagonal elements of the sample covariance matrix (or a deformation thereof) typically are not expected to be too stable out-of-sample. In this regard, instead of using a computed (based on the alpha time series) sample covariance matrix, one can use a much more stable constructed factor model covariance matrix, which we discuss in Section \ref{sub3.1} hereof.

{}To begin with, we will ignore trading costs. Alphas $\alpha_i$ are combined with weights $w_i$. Portfolio P\&L, volatility and Sharpe ratio are given by
\begin{eqnarray}
 &&P = I~\sum_{i=1}^N \alpha_i~w_i\\
 &&R = I~\sqrt{\sum_{i,j=1}^N C_{ij}~w_i~w_j}\\
 &&S = {P \over R}
\end{eqnarray}
where $I$ is the investment level. Any leverage is included in the definition of $\alpha_i$, {\em i.e.}, if a given alpha labeled by index $\ell\in[1,\dots,N]$ before leverage is ${\widetilde\alpha}_\ell$ (this is a raw, unlevered alpha) and the corresponding leverage is $K_\ell:1$, then we define $\alpha_\ell\equiv K_\ell~{\widetilde\alpha}_\ell$. With this definition, the weights satisfy the condition
\begin{equation}\label{w.norm}
 \sum_{i=1}^N \left|w_i\right| = 1
\end{equation}
Here we allow the weights to be negative. This is because here we are primarily interested in the case where the alphas are traded on the same execution platform and trades between alphas are crossed, so one is actually trading the combined alpha. Since generically there are nonzero correlations between different alphas (that is, at least some off-diagonal elements of the correlation matrix $\Psi_{ij}$ are nonzero), the optimal solution can have some negative weights, {\em i.e.}, it is more optimal to trade some alphas reversed.

{}In this paper we will focus on the optimization where one maximizes the Sharpe ratio:
\begin{equation}\label{max.sharpe}
 S \rightarrow \mbox{max}
\end{equation}
We will assume that there are no upper or lower bounds on the weights -- our primary goal here is to set the framework for optimization with linear and nonlinear costs.

{}The solution to (\ref{max.sharpe}) in the absence of costs is given by
\begin{equation}
 w_i = \gamma \sum_{j=1}^N C^{-1}_{ij}\alpha_j
\end{equation}
where $C^{-1}$ is the inverse of $C$, and the normalization coefficient $\gamma$ is determined from (\ref{w.norm}). Without delving into any details, here we simply assume that $C$ is invertible or is made into such ({\em e.g.}, via a deformation -- see, {\em e.g.}, a method discussed in \cite{SpMod} based on \cite{RJ}). We will discuss the case where $C$ is singular in Section \ref{sec.reg}.

{}If $C_{ij}$ is diagonal and we have all $\alpha_i > 0$, then all $w_i$ are also positive. However, when we have nonzero correlations between alphas, some weights can be negative even if all alphas are positive -- a simple example is given in footnote \ref{neg.w}.

\subsection{Linear costs}\label{sub2.1}

{}Linear costs can be modeled by subtracting a linear penalty from the P\&L:
\begin{equation}
 P = I~\sum_{i=1}^N \alpha_i~w_i - L~D
\end{equation}
where $L$ includes all fixed trading costs (SEC fees, exchange fees, broker-dealer fees, {\em etc.}) and linear slippage.\footnote{Here for the sake of simplicity the linear slippage is assumed to be uniform across all alphas. This is not a critical assumption and can be relaxed, {\em e.g.}, by modifying the definition of $L_i$ below. In essence, this assumption is made to simplify the discussion of turnover reduction.\label{fn.lin.cost}} The linear cost assumes no impact, {\em i.e.}, trading does not affect the stock prices. Each alpha is assumed to trade a large number of stocks. Each individual stock has its own contribution to linear cost, which depends on its liquidity, volatility, {\em etc.} When summed over a large number of stocks and a large number of alphas, the linear cost can be modeled (with the caveat mentioned in footnote \ref{fn.lin.cost}) as being proportional to the dollar turnover ({\em i.e.}, the dollar amount traded by the portfolio) $D \equiv I~T$, where $T$ is what we refer to simply as the turnover (so the turnover $T$ is defined as a percentage). Details are relegated to Appendix \ref{app0}, which discussed the relation between, on the one hand, $D$, $T$ and $w_i$ (which are the quantities typically used in the optimization discussions), and, on the other hand, the individual stock prices and shares traded (which are the quantities typically used in the transaction cost discussions).

{}Let $\tau_i$ be the turnovers corresponding to individual alphas $\alpha_i$. If we ignore turnover reduction resulting from combining alphas (or if the internal crossing is switched off), then
\begin{equation}
 T = \sum_{i=1}^N \tau_i~\left|w_i\right|
\end{equation}
However, with internal crossing turnover reduction can be substantial and needs to be taken into account. In \cite{SpMod} we proposed a model of turnover reduction, according to which when the number of alphas $N$ is large, the leading approximation (in the $1/N$ expansion) is given by
\begin{equation}\label{SM}
 T \approx  \rho_* \sum_{i=1}^N \tau_i~\left|w_i\right|
\end{equation}
where $0 < \rho_* \leq 1$ is the turnover reduction coefficient. Let us emphasize that this formula is expected to be a good approximation in the large $N$ limit (so long as the distribution of individual turnovers $\tau_i$ is not skewed) regardless of how $\rho_*$ is modeled. In \cite{SpMod} we also proposed a spectral model for estimating $\rho_*$ based on the correlation matrix $\Psi_{ij}$:
\begin{equation}
 \rho_* \approx {\psi^{(1)}\over{N\sqrt{N}}}~ \left|\sum_{i=1}^N {\widetilde V}^{(1)}_{i}\right|
\end{equation}
where $\psi^{(1)}$ is the largest eigenvalue of $\Psi_{ij}$ and ${\widetilde V}^{(1)}_{i}$ is the corresponding eigenvector normalized such that $\sum_{i=1}^N
\left({\widetilde V}^{(1)}_{i}\right)^2 = 1$.

{}We then have
\begin{equation}\label{L}
 P = I~\sum_{i = 1}^N \left(\alpha_i~w_i - L_i~\left|w_i\right|\right)
\end{equation}
where
\begin{equation}\label{L1}
 L_i \equiv L~\rho_*~\tau_i > 0
\end{equation}
Note that under the rescaling $w_i \rightarrow \zeta w_i$ ($\zeta > 0$) we have $P\rightarrow \zeta P$, $R\rightarrow \zeta R$ and $S = \mbox{inv}$. This allows to recast the Sharpe ratio maximization condition (\ref{max.sharpe}) into the following minimization problem:
\begin{eqnarray}
 && g(w,\lambda) \equiv {\lambda\over 2}\sum_{i,j=1}^N C_{ij}~w_i~ w_j - \sum_{i=1}^N \left(\alpha_i~w_i - L_i~\left|w_i\right|\right) \\
 && g(w,\lambda)\rightarrow \mbox{min}
\end{eqnarray}
where $\lambda > 0$ is a free parameter, which is determined {\em after} the minimization w.r.t. $w_i$ (with $\lambda$ fixed) from the requirement (\ref{w.norm}). If it were not for the modulus in $L_i\left|w_i\right|$, this optimization problem would be solvable in closed form. The modulus complicates things a bit. The problem can still be solved, albeit it requires a finite iterative procedure, {\em i.e.}, the solution (formally) is exact and is obtained after a finite number of iterations.\footnote{\, More precisely, this is the case when the covariance matrix takes a factor model form -- see below.}

\section{Optimization with Linear Costs}\label{sec.opt}

{}Let $J$ and $J^\prime$ be the subsets of the index $i=1,\dots,N$ such that
\begin{eqnarray}
 &&w_i\not= 0,~~~i\in J\\
 &&w_i = 0,~~~i\in J^\prime
\end{eqnarray}
Let
\begin{equation}
 \eta_i \equiv \mbox{sign}\left(w_i\right),~~~i\in J
\end{equation}
Note that, since the modulus has a discontinuous derivative, the minimization equations are not the same as setting first derivatives of $g(w, \lambda)$ w.r.t. $w_i$ to zero. More concretely, first derivatives are well-defined for $i\in J$, but not for $i\in J^\prime$. So, we have the following minimization equations for $w_i$, $i\in J$:
\begin{equation}\label{J1}
 \lambda \sum_{j\in J} C_{ij}~w_j - \alpha_i + L_i~\eta_i  = 0,~~~i\in J
\end{equation}
There are additional conditions for the global minimum\footnote{The global optimum conditions are discussed in Appendix \ref{appA}.} corresponding to the directions $i\in J^\prime$:
\begin{eqnarray}\label{JJ1}
 &&{\lambda\over 2}\sum_{i,j=1}^N C_{ij}~(w_i+\epsilon_i)~(w_j+\epsilon_j) - \sum_{i=1}^N \left(\alpha_i~(w_i+\epsilon_i) - L_i~\left|w_i + \epsilon_i\right|\right) \geq\nonumber\\
 &&\,\,\,\,\,\,\,{\lambda\over 2}\sum_{i,j=1}^N C_{ij}~w_i~w_j - \sum_{i=1}^N \left(\alpha_i~w_i - L_i~\left|w_i\right|\right)
\end{eqnarray}
where $w_i$, $i\in J$ are determined using (\ref{J1}), while $w_i =0$, $i\in J^\prime$. The conditions (\ref{JJ1}) must be satisfied including for arbitrary infinitesimal $\epsilon_i$. Taking into account (\ref{J1}), these conditions can be rewritten as follows:\footnote{\, Since here $\epsilon_i$ are taken to be infinitesimal, these are the conditions for a local minimum. In Appendix \ref{appA} we show that the local minimum we find here is also the global minimum.}
\begin{equation}\label{JJ4}
 \sum_{j\in J^\prime}\left(\lambda \sum_{i\in J} C_{ij}~w_i~\epsilon_j -\alpha_j~\epsilon_j+ L_j \left|\epsilon_j\right| \right)\geq 0
\end{equation}
Since $\epsilon_j$, $j\in J^\prime$ are arbitrary (albeit infinitesimal), this gives the following conditions:
\begin{equation}\label{globalmin}
 \forall j\in J^\prime:~~~\left|\lambda \sum_{i\in J} C_{ij}~w_i - \alpha_j\right| \leq L_j
\end{equation}
These conditions must be satisfied by the solution to (\ref{J1}). The solution that minimizes $g(w, \lambda)$ is given by
\begin{equation}
 w_i = {1\over\lambda}\sum_{j\in J} D_{ij}~\left(\alpha_j - L_j~\eta_j\right),~~~i\in J
\end{equation}
and $D$ is the inverse matrix of the $N(J)\times N(J)$ matrix $C_{ij}$, $i,j\in J$, where $N(J)\equiv\left|J\right|$ is the number of elements of $J$:
\begin{equation}
 \sum_{k\in J} C_{ik}~D_{kj} = \delta_{ij},~~~i,j\in J
\end{equation}
{\em i.e.}, $D$ is {\em not} a restriction of the inverse of the $N\times N$ matrix $C_{ij}$ to $i,j\in J$.

{}Here the following observation is in order. In the above solution, {\em a priori} we do not know i) what the subset $J^\prime$ is and ii) what the values of $\eta_i$ are for $i\in J$. This means that {\em a priori} we have total of $3^N$ possible combinations (including the redundant empty $J$ case), so if we go through this finite set, we will solve the problem exactly. However, $3^N$ is a prohibitively large number for any decent number of alphas, which we in fact assume to be large, so one needs a more clever way of solving the problem.

\subsection{Factor Model}\label{sub3.1}

{}We need to reduce the number of iterations. In this regard, the following observation is useful. Suppose, for a moment, that $C_{ij}$ were diagonal: $C_{ij} = \xi^2_i\delta_{ij}$. Then (\ref{JJ4}) simplifies and we have $w_i = 0$ for $i\in J^\prime$ such that $\left|\alpha_i\right| \leq L_i$, while for $i\in J$ such that $\left|\alpha_i\right| > L_i$ from (\ref{J1}) we have $\eta_i = \mbox{sign}\left(\alpha_i\right)$ and $w_i = \left[\alpha_i - L_i~\mbox{sign}\left(\alpha_i\right)\right]/\lambda \xi^2_i$. {\em I.e.}, in this case we do not need any iterations. This suggests that, if we reduce the ``off-diagonality" of $C_{ij}$, the number of required iterations should also decrease.

{}This can be achieved by considering a factor model for alphas. Just as in the case of a stock multi-factor risk model, instead of $N$ alphas, one deals with $F\ll N$ risk factors and the covariance matrix $C_{ij}$ is replaced by $\Gamma_{ij}$ given by
\begin{eqnarray}\label{Gamma}
 &&\Gamma \equiv \Xi + \Omega~\Phi~\Omega^T\\
 && \Xi_{ij} \equiv \xi_i^2 ~\delta_{ij}
\end{eqnarray}
where $\xi_i$ is the specific risk for each $\alpha_i$; $\Omega_{iA}$ is an $N\times F$ factor loadings matrix; and $\Phi_{AB}$ is the factor covariance matrix, $A,B=1,\dots,F$. {\em I.e.}, the random processes $\Upsilon_i$ corresponding to $N$ alphas are modeled via $N$ random processes $z_i$ (corresponding to specific risk) together with $F$ random processes $f_A$ (corresponding to factor risk):
\begin{eqnarray}
 &&\Upsilon_i = z_i + \sum_{A=1}^F \Omega_{iA}~f_A\\
 &&\left<z_i, z_j\right> = \Xi_{ij}\\
 &&\left<z_i, f_A\right> = 0\\
 &&\left<f_A, f_B\right> = \Phi_{AB}\\
 &&\left<\Upsilon_i, \Upsilon_j\right> = \Gamma_{ij}
\end{eqnarray}
Instead of an $N \times N$ covariance matrix $C_{ij}$ we now have an $F \times F$ covariance matrix $\Phi_{AB}$. So, below we will set
\begin{eqnarray}
 &&C = \Gamma \equiv \Xi + {\widetilde \Omega}~{\widetilde \Omega}^T \\
 &&{\widetilde \Omega} \equiv \Omega~{\widetilde \Phi}\\
 &&{\widetilde \Phi}~{\widetilde \Phi}^T = \Phi
\end{eqnarray}
where ${\widetilde \Phi}_{AB}$ is the Cholesky decomposition of $\Phi_{AB}$, which is assumed to be positive-definite.

{}There are various approaches to constructing factor models for alpha streams. Here we simply assume a factor model form for the covariance matrix without delving into details of how it is constructed.\footnote{\, A more detailed discussion of factor models for alpha streams will appear in a forthcoming paper.} Let us briefly mention one evident possibility: one can use the first $F$ principal components of the covariance matrix as the factor loadings matrix. One then needs to construct specific risk and factor covariance matrix (which in itself is nontrivial). This is essentially the APT approach.

\subsection{Optimization with Factor Model}

{}In the factor-model framework, the optimization problem reduces to solving an $F$-dimensional system as follows. First, let
\begin{equation}\label{v}
 v_A \equiv \sum_{i=1}^N~w_i~{\widetilde \Omega}_{iA} = \sum_{i\in J}~w_i~{\widetilde \Omega}_{iA},~~~A=1,\dots,F
\end{equation}
Then from (\ref{J1}) we have
\begin{equation}\label{wv}
 w_i = {1\over \lambda \xi_i^2}~\left(\alpha_i - L_i~\eta_i - \lambda \sum_{A=1}^F {\widetilde \Omega}_{iA}~v_A\right),~~~i\in J
\end{equation}
Recalling that we have
\begin{equation}\label{w-eta}
 w_i~\eta_i > 0,~~~i\in J
\end{equation}
we get
\begin{eqnarray}
 &&\eta_i = \mbox{sign}\left(\alpha_i - \lambda \sum_{A=1}^F {\widetilde \Omega}_{iA}~v_A\right),~~~i\in J\label{eta1}\\
 &&\forall i\in J:~~~\left|\alpha_i - \lambda \sum_{A=1}^F {\widetilde \Omega}_{iA}~v_A\right| > L_i \label{eta2}\\
 &&\forall i\in J^\prime:~~~\left|\alpha_i - \lambda \sum_{A=1}^F {\widetilde \Omega}_{iA}~v_A\right| \leq L_i\label{eta3}
\end{eqnarray}
where (\ref{eta2}) follows from (\ref{wv}) and (\ref{w-eta}). The last two inequalities define $J$ and $J^\prime$ in terms of $F$ unknowns $v_A$.

{}Substituting (\ref{wv}) into (\ref{v}), we get the following system of $F$ equations for $F$ unknowns $v_A$:
\begin{equation}
 \sum_{B=1}^F Q_{AB}~v_B = a_A
\end{equation}
where
\begin{eqnarray}
 &&Q_{AB} \equiv \delta_{AB} + \sum_{i\in J} {{{\widetilde \Omega}_{iA}~{\widetilde \Omega}_{iB}} \over {\xi_i^2}}\\
 &&a_A \equiv {1\over\lambda} \sum_{i\in J} {{{\widetilde \Omega}_{iA}} \over {\xi_i^2}}\left[\alpha_i - L_i~\eta_i\right]
\end{eqnarray}
so we have
\begin{equation}\label{Qv}
 v_A = \sum_{B=1}^F Q^{-1}_{AB}~a_B
\end{equation}
where $Q^{-1}$ is the inverse of $Q$.

{}Note that (\ref{Qv}) solves for $v_A$ given $\eta_i$, $J$ and $J^\prime$. On the other hand, (\ref{eta1}), (\ref{eta2}) and (\ref{eta3}) determine $\eta_i$, $J$ and $J^\prime$ in terms of $v_A$. The entire system can then be solved iteratively.

{}An algorithm for an iterative procedure for solving the system (\ref{eta1}), (\ref{eta2}), (\ref{eta3}) and (\ref{Qv}) is relegated to Appendix \ref{appB}. Let us emphasize that the iterative procedure is finite, {\em i.e.}, it converges in a finite number of iterations.

\section{Impact in Weight Optimization}\label{sec.impact}

{}Next, let us discuss the effect of impact, {\em i.e.}, nonlinear costs, on weight optimization. Generally, introducing nonlinear impact makes the weight optimization problem computationally more challenging and requires introduction of approximation methods.

{}One way of modeling trading costs is to introduce linear and nonlinear terms:
\begin{equation}
 P = I~\sum_{i=1}^N \alpha_i~w_i - L~D - {1\over n}~Q~D^n
\end{equation}
where $D = I~T$ is the dollar amount traded, $T$ is the turnover, and $Q$ and $n>1$ are model-dependent (and can be measured empirically). If we model turnover using (\ref{SM}), then we have
\begin{equation}\label{impact}
 P = I~\sum_{i=1}^N \left(\alpha_i~w_i - L_i~\left|w_i\right|\right) - {{\widetilde Q}\over n}\left[\sum_{i=1}^N \tau_i~\left|w_i\right|\right]^n
\end{equation}
where the modulus accounts for the possibility of some $w_i$ being negative, and ${\widetilde Q}$ is defined as follows
\begin{equation}
 {\widetilde Q}\equiv Q~(I~\rho_*)^n
\end{equation}
For general fractional $n$, which would have to be measured empirically, the weight optimization problem would have to be solved numerically. Sometimes $n$ is assumed to be 3/2. Here we keep it arbitrary.

{}First, note that if individual turnovers $\tau_i\equiv\tau$ are identical, then the nonlinear cost contribution into $P$ is independent of $w_i$ as we have (\ref{w.norm}). In this case, it simply shifts $P$ by a constant and the problem can be solved exactly as in the previous section.\footnote{\, In fact, in this case the contribution of the linear cost also shifts $P$ by a constant.} If $\tau_i$ are not all identical, then we need to solve the following problem:
\begin{eqnarray}
 &&g(w, \lambda) \equiv {1\over 2}~\sum_{i,j=1}^N C_{ij}~w_i~w_j - \sum_{i=1}^N \left(\alpha_i~w_i - L_i~\left|w_i\right|\right) +{{\widetilde Q}^\prime \over n} \left[\sum_{i=1}^N \tau_i~\left|w_i\right|\right]^n \\
 &&g(w, \mu, {\widetilde \mu}) \rightarrow \mbox{min}
\end{eqnarray}
where
\begin{equation}
 {\widetilde Q}^\prime \equiv {{\widetilde Q}\over I}
\end{equation}
Here one can use successive iterations to deal with the nonlinear term and various stability issues associated with convergence must be addressed. A simpler approach is to note that the key role of the nonlinear term is to model portfolio capacity\footnote{\label{capacity}\, By this we mean the value of the investment level $I = I_*$ for which the P\&L $P_{\rm{\scriptstyle{opt}}}(I)$ is maximized, where for any given $I$ P\&L $P_{\rm{\scriptstyle{opt}}}(I)$ is computed for the optimized weights $w_i$. When only linear cost is present, capacity is unbounded. When nonlinear cost is included, $I_*$ is finite.} via its dependence on $I$, not its detailed structure in terms of individual alphas. In this regard, the following approximation is a reasonable way of simplifying the problem.

{}Let
\begin{eqnarray}
 &&{\overline \tau}\equiv {1\over N}\sum_{i=1} \tau_i\\
 &&{\widetilde \tau}_i\equiv \tau_i - {\overline \tau}
\end{eqnarray}
If the distribution of ${\widetilde \tau}_i$ has a small standard deviation, then we can use the following approximation (where we are using (\ref{w.norm})):
\begin{equation}
 \left[\sum_{i=1}^N \tau_i~\left|w_i\right|\right]^n\approx {\overline \tau}^n + n~{\overline \tau}^{n-1}~\sum_{i=1}^N {\widetilde \tau}_i~\left|w_i\right|
\end{equation}
The objective function can be rewritten as (modulo an immaterial constant term)
\begin{equation}
 g(w, \lambda) \approx {1\over 2}~\sum_{i,j=1}^N C_{ij}~w_i~w_j - \sum_{i=1}^N \left(\alpha_i~w_i - {\widetilde L}_i~\left|w_i\right|\right)
\end{equation}
where
\begin{equation}
 {\widetilde L}_i \equiv L_i + {\widetilde Q}^\prime~{\overline \tau}^{n-1}~\tau_i = L_i + Q~\rho_*^n~I^{n-1}~{\overline \tau}^{n-1}~\tau_i \label{Leff}
\end{equation}
{\em I.e.}, in this approximation the effect of the nonlinear term reduces to increasing the linear slippage, and this problem we can solve as in the previous section. Note, however, that the ``effective" linear cost ${\widetilde L}_i$ now depends on the investment level $I$ via (\ref{Leff}), which now controls capacity. Thus, for $I$ such that
\begin{equation}
 \forall i=1,\dots,N:~~~{\widetilde L}_i \geq \left|\alpha_i\right|
\end{equation}
the P\&L cannot be positive, so the capacity $I_*$ is finite (see footnote \ref{capacity}).

\section{Regression as Limit of Optimization}\label{sec.reg}

{}Let us go back to optimization without costs. The Sharpe ratio is maximized by
\begin{equation}\label{opt}
 w_i = \gamma~\sum_{j=1}^N C^{-1}_{ij}~\alpha_j
\end{equation}
where $\gamma$ is a normalization constant.

{}Let $C_{ij}$ have a factor model form:
\begin{equation}
 C_{ij} = v_i~\delta_{ij} + \sum_{A = 1}^K~\Lambda_{iA}~\Lambda_{jA}
\end{equation}
where $v_i$ is specific variance, and $\Lambda_{iA}$, $A=1,\dots,K$ is the factor loadings matrix in the basis where the factor covariance matrix is the identity matrix.\footnote{\, {\em I.e.}, the factor covariance matrix is absorbed into the definition of the factor loadings matrix.}

{}We have
\begin{equation}
 w_i = {\gamma\over v_i}~\left(\alpha_i - \sum_{j = 1}^N {\alpha_j \over v_j}~\sum_{A,B = 1}^K \Lambda_{iA}~\Lambda_{jB}~Q^{-1}_{AB} \right)
\end{equation}
where $Q^{-1}_{AB}$ is the inverse of
\begin{equation}
 Q_{AB}\equiv \delta_{AB} + \sum_{\ell = 1}^N {1\over v_\ell}~\Lambda_{\ell A}~\Lambda_{\ell B}
\end{equation}
Note that for $N=1$ and $K=1$ we have
\begin{equation}
 w_1 = {{\gamma~\alpha_1}\over{v_1 + \Lambda_{11}^2}}
\end{equation}
which reproduces (\ref{opt}).

\subsection{Regression Limit}

{}Let
\begin{equation}
 v_i \equiv \zeta~{\widetilde v}_i
\end{equation}
Consider the following limit:
\begin{eqnarray}
 &&\zeta \rightarrow 0\\
 &&\gamma\rightarrow 0\\
 &&{\gamma\over \zeta}\equiv {\widetilde \gamma} = \mbox{fixed}\\
 &&{\widetilde v}_i = \mbox{fixed}
\end{eqnarray}
In this limit we have
\begin{equation}
 w_i = {{\widetilde \gamma}\over {\widetilde v}_i}~\left(\alpha_i - \sum_{j = 1}^N {\alpha_j \over {\widetilde v}_j}~\sum_{A,B = 1}^K \Lambda_{iA}~\Lambda_{jB}~{\widetilde Q}^{-1}_{AB} \right) \equiv {{\widetilde \gamma}\over {\widetilde v}_i}~\varepsilon_i
\end{equation}
where
${\widetilde Q}^{-1}_{AB}$ is the inverse of
\begin{equation}
 {\widetilde Q}_{AB}\equiv \sum_{\ell = 1}^N {1\over {\widetilde v}_\ell}~\Lambda_{\ell A}~\Lambda_{\ell B}
\end{equation}
Note that
\begin{equation}
 \sum_{i=1}^N w_i~\Lambda_{iC} \equiv 0,~~~C=1,\dots, K
\end{equation}
In fact, $\varepsilon_i$ are the residuals of a weighted regression (with weights $1/{\widetilde v}_i$) of $\alpha_i$ over $\Lambda_{iA}$ (without intercept). If all weights are identical ${\widetilde v}_i\equiv {\widetilde v}$, then we have an equally-weighted regression:
\begin{equation}
 \alpha_i = \sum_{A=1}^K \Lambda_{iA}~\eta_A + \varepsilon_i
\end{equation}
where $\eta_A$ are the regression coefficients (in matrix notation): $\eta = \left(\Lambda^T~\Lambda\right)^{-1}~\Lambda^T~\alpha$.

\subsection{Regression Limit with Costs}

{}We can take a similar limit in the solution of Section 3 with costs. In this limit we have
\begin{eqnarray}
 &&\xi^2_i \equiv \zeta~{\widetilde \xi}^2_i\\
 &&\lambda \equiv {\widetilde\lambda}/\zeta\\
 &&\zeta\rightarrow 0\\
 &&{\widetilde \xi}^2_i = \mbox{fixed}\\
 &&{\widetilde \lambda} = \mbox{fixed}
\end{eqnarray}
In this limit (\ref{wv}) reduces to
\begin{equation}\label{wv1}
 w_i = {\varepsilon_i\over{\widetilde\lambda}~{\widetilde\xi}^2_i}
\end{equation}
where $\varepsilon_i$ are the residuals of a weighted regression (with weights $1/{\widetilde\xi}^2_i$) of $\alpha_i - L_i~\eta_i$ over ${\widetilde \Omega}_{iA}$ (without intercept). We can use (\ref{wv1}) (instead of (\ref{wv})) in the iterative procedure discussed at the end of Section 3, which now defines ``Regression with Linear Costs" (as opposed to optimization with linear costs) and can be useful in cases where the full factor model is not known, but factor loadings ${\widetilde \Omega}_{iA}$ can be constructed. An example of this is when the number of observations $(M+1)$ for alphas is small ($M\ll N$), so the covariance matrix $C_{ij}$ is singular and has only $M$ non-vanishing eigenvalues $e_A$. In this case one can use, {\em e.g.}, the first $M$ principal components $P_{iA}$ (corresponding to the nonzero eigenvalues $e_A$) to construct factor loadings via ${\widetilde\Omega}_{iA} = \sqrt{e_A}P_{iA}$, and for ${\widetilde\xi}_i^2$ one can use, {\em e.g.}, $C_{ii}$ (which are all positive).\footnote{\, Note that for the regression one can actually set ${\widetilde\Omega}_{iA} = P_{iA}$ as any transformation of the form ${\widetilde\Omega}\rightarrow {\widetilde\Omega}~Z$, where $Z$ is an arbitrary nonsingular $M\times M$ matrix, does not change the regression residuals (albeit it affects the regression coefficients).} Finally, note that we can also consider regression with linear and nonlinear costs with the latter treated using the approximation discussed in Section \ref{sec.impact}.

\appendix

\section{Linear Costs}\label{app0}

{}In this appendix we discuss linear costs in more detail, starting from linear costs for underlying individual stocks, which we discuss in terms of the individual stock prices $P_A$ and the corresponding volumes traded $Q_{iA}$. Here the index $A=1,\dots,N_S$ labels stocks, where $N_S$ is the total number of stocks traded. As before, $i=1,\dots,N$, where $N$ is the number of alphas. Then $Q_{iA}$ is the volume ({\em i.e.}, the number of shares) for the stock labeled by $A$ traded by $\alpha_i$. Here volumes $Q_{iA}$ are unsigned quantities, {\em i.e.}, $Q_{iA} \geq 0$ both for buys and sells. Let $L_{iA}$ be the per-share linear cost of trading the stock labeled by $A$ by $\alpha_i$. First, let us assume that there is no internal crossing. Then the total linear cost of trading all stocks by all alphas is given by
\begin{equation}
 C_{\rm{\scriptstyle{lin}}} = \sum_{i=1}^N \sum_{A=1}^{N_S} L_{iA}~Q_{iA}
\end{equation}
This equation, however, is not practical for the purpose of weight optimization. We need to make some simplifying assumptions, so we can express $C_{\rm{\scriptstyle{lin}}}$ in terms of the weights $w_i$. The two simplifying assumptions are as follows. First, we assume that $L_{iA}$ is independent of the $i$ index, {\em i.e.}, the cost of trading the stock labeled by $A$ is independent of which alpha is trading it. This assumption need not hold in the most general case. However, when the number of stocks $N_S$ is large and the number of alphas $N$ is large, this is expected to be a reasonable approximation, which can be thought of as setting $L_{iA}$ to their mean value (as averaged over all alphas)
\begin{equation}
 L_{iA} \approx L_A\equiv{1\over N} \sum_{i=1}^N L_{iA}
\end{equation}
Second, in optimization one deals with {\em dollar} holdings, not share holdings -- thus, the total (meaning, long {\em plus} short) dollar holding for each alpha is $H_i \equiv I~|w_i|$. On the other hand, we can write
\begin{equation}
 H_i = \sum_{A=1}^{N_S} P_A~S_{iA}
\end{equation}
where $S_{iA}$ is the absolute value of shares held by $\alpha_i$ in the stock labeled by $A$. Similarly, to tackle the optimization problem, trading costs should also be given in terms of traded {\em dollar} amounts. This is achieved by assuming that $L_A$ is proportional to the prices $P_A$, {\em i.e.}, $L_A \approx L~P_A$, where $L$ is independent of $A$. We then have
\begin{equation}\label{C.lin}
 C_{\rm{\scriptstyle{lin}}}\approx L \sum_{i=1}^N \sum_{A=1}^{N_S} P_A~Q_{iA} = L~D
\end{equation}
The second equality follows from the definition of the portfolio dollar turnover $D = \sum_{i=1}^N D_i$, where $D_i = \sum_{A=1}^{N_S} P_A~Q_{iA}$ are individual dollar turnovers in the absence of internal crossing. (We discuss turnover reduction in the presence of internal crossing in Section \ref{sub2.1}.) Note that $D_i = D~|w_i|$, and $T = D/I$. Furthermore, the meaning of (\ref{C.lin}) is that the linear cost approximately is a fixed fraction of the dollar amount traded. This is expected to be a reasonable approximation when linear slippage has a dominant contribution into the linear cost -- linear slippage for an individual stock is roughly proportional to an average bid-ask spread, which on average scales linearly with the stock price, so when the linear cost is summed over a large number of stocks and a large number of alphas, we arrive at the above approximation.

\section{Conditions for Global Minimum}\label{appA}

{}In Section \ref{sec.opt} we gave the conditions for the global minimum:
\begin{eqnarray}\label{AJJ1}
 &&{\lambda\over 2}\sum_{i,j=1}^N C_{ij}~(w_i+\epsilon_i)~(w_j+\epsilon_j) - \sum_{i=1}^N \left(\alpha_i~(w_i+\epsilon_i) - L_i~\left|w_i + \epsilon_i\right|\right) \geq\nonumber\\
 &&\,\,\,\,\,\,\,{\lambda\over 2}\sum_{i,j=1}^N C_{ij}~w_i~w_j - \sum_{i=1}^N \left(\alpha_i~w_i - L_i~\left|w_i\right|\right)
\end{eqnarray}
where $w_i$, $i\in J$ are determined using (\ref{J1}), while $w_i =0$, $i\in J^\prime$, and $\epsilon_i$ are arbitrary. In Section \ref{sec.opt} we discussed these conditions for arbitrary infinitesimal $\epsilon_i$, which gave the conditions for a local minimum. Here we discuss the above conditions for non-infinitesimal $\epsilon_i$. Taking into account (\ref{J1}), we have
\begin{eqnarray}
 &&{\lambda\over 2}\sum_{i,j=1}^N C_{ij}~\epsilon_i~\epsilon_j + \sum_{j\in J^\prime}\left(\lambda \sum_{i\in J} C_{ij}~w_i~\epsilon_j -\alpha_j~\epsilon_j+ L_j \left|\epsilon_j\right| \right) + \nonumber\\
 &&\,\,\,\,\,\,\,\sum_{i\in J} L_i~\left(\left|w_i + \epsilon_i\right| - \left|w_i\right| - \eta_i~\epsilon_i\right) \geq  0\label{AAJJ2}
\end{eqnarray}
The first term is manifestly positive semi-definite as $C_{ij}$ is positive-definite, the second term is positive semi-definite due to (\ref{globalmin}) which implies (\ref{JJ4}), while the third term is manifestly positive semi-definite as $\eta_i = \mbox{sign}(w_i)$. So, the local minimum we found in Section \ref{sec.opt} is also the global minimum. This is because all $L_i > 0$.

\section{Iterative Procedure}\label{appB}

{}At the initial iteration one takes $J^{(0)}=\{1,\dots,N\}$, so that $J^{\prime(0)}$ is empty, and
\begin{equation}
 \eta^{(0)}_i = \pm 1,~~~i=1,\dots,N
\end{equation}
While {\em a priori} the values of $\eta^{(0)}_i$ can be arbitrary, unless $F\ll N$, in some cases one might encounter convergence speed issues. However, if one chooses
\begin{equation}
 \eta^{(0)}_i = \mbox{sign}(\alpha_i),~~~i=1,\dots,N
\end{equation}
then the iterative procedure generally is expected to converge rather fast. Furthermore, note that the solution is actually exact, {\em i.e.}, the convergence criteria are given by (recall from Appendix \ref{appA} that this produces the global optimum)
\begin{eqnarray}
 &&J^{(s + 1)} = J^{(s)}\\
 &&\forall i\in J^{(s+1)}:~~~\eta^{(s+1)}_i = \eta^{(s)}_i\\
 &&\forall A\in \{1,\dots,F\}:~~~v^{(s + 1)}_A = v^{(s)}_A
\end{eqnarray}
where $s$ and $s+1$ label successive iterations.\footnote{\, The first two of these criteria are based on discrete quantities and are unaffected by computational (machine) precision effects, while the last criterion is based on continuous quantities and in practice is understood as satisfied within computational (machine) precision or preset tolerance.} Put differently, the iterative procedure is finite -- it converges in a finite number of iterations. Finally, note that $w_i$ for $i\in J$ are given by (\ref{wv}), while $w_i = 0$ for $i\in J^\prime$.

{}Here the following remark is in order. Because the alphas $\alpha_i$, $i\in J^\prime$ are no longer traded, we can drop such alphas, if any, recompute $\rho_*$ in (\ref{L1}) using the corresponding correlation matrix $\Psi^\prime_{ij} \equiv \Psi_{ij}\left.\right|_{i,j\in J}$, recompute $w_i$ using such $\rho_*$ and repeat this procedure until the subset $J$ based on which $\rho_*$ is computed is the same as the subset for which $w_i\not=0$, where $w_i$ are computed based on such $\rho_*$.\footnote{\, When $N$ is large, this procedure is stable and convergent as $\rho_*$ does not change much with $N$ (see \cite{SpMod}).}


\begin{thebibliography}{}

\bibitem{HF1} T. Schneeweis, R. Spurgin, and D. McCarthy,
``Survivor Bias in Commodity Trading Advisor Performance",
J. Futures Markets, 1996, 16(7), 757-772.

\bibitem{HF2} C. Ackerman, R. McEnally and D. Revenscraft,
``The Performance of Hedge Funds: Risk, Return and Incentives",
Journal of Finance, 1999, 54(3), 833-874.

\bibitem{HF3} S.J. Brown, W. Goetzmann and R.G. Ibbotson,
``Offshore Hedge Funds: Survival and Performance, 1989-1995",
Journal of Business, 1999, 72(1), 91-117.

\bibitem{HF4} F.R. Edwards and J. Liew,
``Managed Commodity Funds",
Journal of Futures Markets, 1999, 19(4), 377-411.

\bibitem{HF5} F.R. Edwards and J. Liew,
``Hedge Funds versus Managed Futures as Asset Classes",
Journal of Derivatives, 1999, 6(4), 45-64.

\bibitem{HF6} W. Fung and D. Hsieh,
``A Primer on Hedge Funds",
Journal of Empirical Finance, 1999, 6(3), 309-331.

\bibitem{HF7} B. Liang,
``On the Performance of Hedge Funds",
Financial Analysts Journal, 1999, 55(4), 72-85.

\bibitem{HF8} V. Agarwal and N.Y. Naik,
``On Taking the ``Alternative" Route: The Risks, Rewards, and Performance Persistence of Hedge Funds",
Journal of Alternative Investments, 2000, 2(4), 6-23.

\bibitem{HF9} V. Agarwal and N.Y. Naik,
``Multi-Period Performance Persistence Analysis of Hedge Funds Source",
Journal of Financial and Quantitative Analysis, 2000, 35(3), 327-342.

\bibitem{HF10} W. Fung and D. Hsieh,
``Performance Characteristics of Hedge Funds and Commodity Funds: Natural vs. Spurious Biases",
Journal of Financial and Quantitative Analysis, 2000, 35(3), 291-307.

\bibitem{HF11} B. Liang,
``Hedge Funds: The Living and the Dead",
Journal of Financial and Quantitative Analysis, 2000, 35(3), 309-326.

\bibitem{HF12} C.S. Asness, R.J. Krail, and J.M. Liew,
``Do Hedge Funds Hedge?",
Journal of Portfolio Management, 2001, 28(1), 6-19.

\bibitem{HF13} F.R. Edwards and M.O. Caglayan,
``Hedge Fund and Commodity Fund Investments in Bull and Bear Markets",
Journal of Portfolio Management, 2001, 27(4), 97-108.

\bibitem{HF14} W. Fung and D. Hsieh,
``The Risk in Hedge Fund Strategies: Theory and Evidence from Trend Followers",
Review of Financial Studies, 2001, 14(2), 313-341.

\bibitem{HF15} B. Liang,
``Hedge Fund Performance: 1990-1999",
Financial Analysts Journal, 2001, 57(1), 11-18.

\bibitem{HF16} A.W. Lo,
``Risk Management For Hedge Funds: Introduction and Overview",
Financial Analysis Journal, 2001, 57(6), 16-33.

\bibitem{HF17} C. Brooks and H.M. Kat,
``The Statistical Properties of Hedge Fund Index Returns and Their Implications for Investors",
Journal of Alternative Investments, 2002, 5(2), 26-44.

\bibitem{HF18} D.-L. Kao,
``Battle for Alphas: Hedge Funds versus Long-Only Portfolios",
Financial Analysts Journal, 2002, 58(2), 16-36.

\bibitem{HF19} G. Amin and H. Kat,
``Stocks, Bonds and Hedge Funds: Not a Free Lunch!",
Journal of Portfolio Management, 2003, 29(4), 113-120.

\bibitem{HF20} N. Chan, M. Getmansky, S.M. Haas and A.W. Lo,
``Systemic Risk and Hedge Funds",
published in: Carey, M. and Stulz, R.M., eds.,
``The Risks of Financial Institutions" (University of Chicago Press, 2006), Chapter 6, 235-338.

\bibitem{PO1} H. Markowitz,
``Portfolio selection",
Journal of Finance, 1952, 7(1), 77-91.

\bibitem{PO2} A. Charnes and W.W. Cooper,
``Programming with linear fractional functionals",
Naval Research Logistics Quarterly, 1962, 9(3-4), 181-186.

\bibitem{PO3} W.F. Sharpe,
``Mutual fund performance",
Journal of Business, 1966, 39(1), 119-138.

\bibitem{PO4} R.C. Merton,
``Lifetime portfolio selection under uncertainty: the continuous time case",
The Review of Economics and Statistics, 1969, 51(3), 247-257.

\bibitem{PO5} S. Schaible,
``Parameter-free convex equivalent and dual programs of fractional programming problems",
Zeitschrift f\"ur Operations Research, 1974, 18(5), 187-196.

\bibitem{PO6} M. Magill and G. Constantinides,
``Portfolio selection with transactions costs",
J. Econom. Theory, 1976, 13(2), 245-263.

\bibitem{PO7} A.F. Perold,
``Large-scale portfolio optimization",
Management Science, 1984, 30(10), 1143-1160.

\bibitem{PO8} M. Davis and A. Norman,
``Portfolio selection with transaction costs",
Math. Oper. Res., 1990, 15(4), 676-713.

\bibitem{PO9} B. Dumas and E. Luciano,
``An exact solution to a dynamic portfolio choice problem under transaction costs",
The Journal of Finance, 1991, 46(2), 577-595.

\bibitem{PO10} C. J. Adcock and N. Meade
``A simple algorithm to incorporate transactions costs in quadratic optimization",
European Journal of Operational Research, 1994, 79(1), 85-94.

\bibitem{PO11} S. Shreve and H.M. Soner,
``Optimal investment and consumption with transaction costs",
Ann. Appl. Probab., 1994, 4(3), 609-692.

\bibitem{PO12} D. Bienstock,
``Computational study of a family of mixed-integer quadratic programming problems",
Mathematical Programming, 1996, 74(2), 121-140.

\bibitem{PO13} J. Cvitani\'{c} and I. Karatzas,
``Hedging and portfolio optimization under transaction costs: a martingale approach",
Math. Finance, 1996, 6(2), 133-165.

\bibitem{PO14} A. Yoshimoto,
``The mean-variance approach to portfolio optimization subject to transaction costs",
J. Operations Research Soc. of Japan, 1996, 39(1), 99-117.

\bibitem{PO15} C. Atkinson, S.R. Pliska and P. Wilmott,
``Portfolio management with transaction costs",
Proc. Roy. Soc. London Ser. A, 1997, 453(1958), 551-562.

\bibitem{PO16} D. Bertsimas, C. Darnell and R. Soucy,
``Portfolio construction through mixed-integer programming at Grantham, Mayo, Van Otterloo and Company",
Interfaces, 1999, 29(1), 49-66.

\bibitem{PO17} A. Cadenillas and S. R. Pliska,
``Optimal trading of a security when there are taxes and transaction costs",
Finance and Stochastics, 1999, 3(2), 137-165.

\bibitem{PO18} T.-J. Chang, N. Meade, J.E. Beasley and Y.M. Sharaiha,
``Heuristics for cardinality constrained portfolio optimisation",
Computers and Operations Research, 2000, 27(13), 1271-1302.

\bibitem{PO19} H. Kellerer, R. Mansini and M.G. Speranza,
``Selecting portfolios with fixed costs and minimum transaction lots",
Annals of Operations Research, 2000, 99(1-4), 287-304.

\bibitem{PO20} R.T. Rockafellar and S. Uryasev,
``Optimization of conditional value-at-risk",
Journal of Risk, 2000, 2(3), 21-41.

\bibitem{PO21} J. Gondzio and R. Kouwenberg,
``High-performance computing for asset-liability management",
Operations Research, 2001, 49(6), 879-891.

\bibitem{PO22} H. Konno and A. Wijayanayake,
``Portfolio optimization problem under concave transaction costs and minimal transaction unit constraints",
Mathematical Programming, 2001, 89(2), 233-250.

\bibitem{PO23} S. Mokkhavesa and C. Atkinson,
``Perturbation solution of optimal portfolio theory with transaction costs for any utility function",
IMA J. Manag. Math., 2002, 13(2), 131-151.

\bibitem{PO24} O.L.V. Costa and A.C. Paiva,
``Robust portfolio selection using linear-matrix inequalities",
Journal of Economic Dynamics and Control, 2002, 26(6), 889-909.

\bibitem{PO25} F. Alizadeh and D. Goldfarb,
``Second-order cone programming",
Mathematical Programming, 2003, 95(1), 3-51.

\bibitem{PO26} M.J. Best and J. Hlouskova,
``Portfolio selection and transactions costs",
Computational Optimization and Applications, 2003, 24(1), 95-116.

\bibitem{PO27} K. Jane\v{c}ek and S. Shreve,
``Asymptotic analysis for optimal investment and consumption with transaction costs",
Finance Stoch., 2004, 8(2), 181-206.

\bibitem{PO28} M.S. Lobo, M. Fazel and S. Boyd,
``Portfolio optimization with linear and fixed transaction costs",
Annals of Operations Research, 2007, 152(1), 341-365.

\bibitem{PO29} R. Zagst and D. Kalin,
``Portfolio optimization under liquidity costs",
International Journal of Pure and Applied Mathematics, 2007, 39(2), 217-233.

\bibitem{PO30} M. Potaptchik, L. Tun\c{c}el and H. Wolkowicz,
``Large scale portfolio optimization with piecewise linear transaction costs",
Optimization Methods and Software, 2008, 23(6), 929-952.

\bibitem{PO31} E. Moro, J. Vicente, L.G. Moyano, A. Gerig, J.D. Farmer, G. Vaglica, F. Lillo and R.N. Mantegna,
``Market impact and trading profile of hidden orders in stock markets",
Physical Review E, 2009, 80, 066102.

\bibitem{PO32} J. Goodman and D.N. Ostrov,
``Balancing small transaction costs with loss of optimal allocation in dynamic stock trading strategies",
SIAM J. Appl. Math., 2010, 70(6), 1977-1998.

\bibitem{PO33} M. Bichuch,
``Asymptotic analysis for optimal investment in finite time with transaction costs",
SIAM J. Financial Math., 2012, 3(1), 433-458.

\bibitem{PO34} J.E. Mitchell and S. Braun,
``Rebalancing an investment portfolio in the presence of convex transaction costs, including market impact costs",
Optimization Methods and Software, 2013, 28(3), 523-542.

\bibitem{PO35} H. Soner and N. Touzi,
``Homogenization and asymptotics for small transaction costs",
SIAM Journal on Control and Optimization, 2013, 51(4), 2893-2921.

\bibitem{OD} Z. Kakushadze and J.K.-S. Liew,
``Is It Possible to OD on Alpha?", SSRN Working Paper, http://ssrn.com/abstract=2419415 (April 2, 2014); arXiv:1404.0746.

\bibitem{SpMod} Z. Kakushadze, ``Spectral Model of Turnover Reduction",
SSRN Working Paper, http://ssrn.com/abstract=2427049 (April 20, 2014); arXiv:1404.5050.

\bibitem{RJ} R. Rebonato and P. J\"ackel,
``The most general methodology to create a valid correlation matrix for risk management and option pricing purposes" (1999), http://ssrn.com/abstract=1969689 (December 7, 2011).

\end{thebibliography}
\end{document}